\documentclass[aps,prd,showkeys,nofootinbib,superscriptaddress]{revtex4-2}
\usepackage{graphicx}
\usepackage[sort&compress]{natbib}
\usepackage{latexsym}
\usepackage{epsfig}
\usepackage{hyperref}
\usepackage{color}
\usepackage{amssymb}
\definecolor{darkgreen}{rgb}{0,0.35,0}
\newcommand{\rc}{\textcolor{red}}

\begin{document}
\newcommand{\eg}{{\it e.g.}}
\newcommand{\etal}{{\it et. al.}}
\newcommand{\ie}{{\it i.e.}}
\newcommand{\be}{\begin{equation}}
\newcommand{\dd}{\displaystyle}
\newcommand{\ee}{\end{equation}}
\newcommand{\bea}{\begin{eqnarray}}
\newcommand{\eea}{\end{eqnarray}}
\newcommand{\bef}{\begin{figure}}
\newcommand{\eef}{\end{figure}}
\newcommand{\bce}{\begin{center}}
\newcommand{\ece}{\end{center}}
\def\lsim{\mathrel{\rlap{\lower4pt\hbox{\hskip1pt$\sim$}}
    \raise1pt\hbox{$<$}}}         
\def\gsim{\mathrel{\rlap{\lower4pt\hbox{\hskip1pt$\sim$}}
    \raise1pt\hbox{$>$}}}         

\title{Quantum black holes, partition of integers and self-similarity}
\author{P. Castorina$^{(a,b)}$, A Iorio $^{(a)}$ and L. Smaldone$^{(c)}$}
\affiliation{$^{(a)}$  Institute of Particle and Nuclear Physics, Faculty  of  Mathematics  and  Physics, Charles  University, V  Hole\v{s}ovi\v{c}k\'{a}ch  2, 18000  Praha  8,  Czech  Republic \\
$^{(b)}$ INFN Sezione di Catania, Via Santa Sofia 64, I-95123 Catania, Italy  \\
$^{(c)}$ Faculty of Physics, Institute of Theoretical Physics, University of Warsaw, Ul. Pasteura 5, 02-093, Warsaw, Poland}

\email{castorina@ct.infn.it}
\email{iorio@mff.cuni.cz}
\email{luca.smaldone@fuw.edu.pl}

\begin{abstract}
We take the view that the area of a black hole's event horizon is quantized, $A = l_P^2 \, (4 \ln 2) \, N$, and the associated degrees of freedom are finite in number and of fermionic nature. We then investigate general aspects of the entropy, $S_{BH}$, our main focus being black-hole self-similarity. We first find a two-to-one map between the black hole's configurations and the ordered partitions of the integer $N$. Hence we construct from there a composition law between the sub-parts making the whole configuration space. This gives meaning to black hole self-similarity, entirely within a single description, as a phenomenon stemming from the well known self-similarity of the ordered partitions of $N$. Finally, we compare the above to the well-known results on the subleading (quantum) corrections, that necessarily require different (quantum) statistical weights for the various configurations.
\end{abstract}


 \maketitle

\section{Introduction}

In this paper we elaborate on the view that the entropy of a black hole, $S_{BH}$, stems from the quantization of the area of the BH's event horizon \cite{mukhanov,bekenstein2, bekenstein3}
\be
A  \ = \  \alpha \, N  \, l_P^2 \,,
\ee
where $N \in \mathbb{N}$, $l_P$ is the Planck length, and the ``it from bit'' \cite{wheelerit} choice for the proportionality factor, $\alpha = 4 \ln 2$, allows for a spin-system description: one bit of information, 0 or 1, $\uparrow$ or $\downarrow$, etc, per given Planck cell. In other words
\be \label{ClassicalEntropy}
S_{BH} \ = \ \frac{A }{4 \, l_P^2} \ = \ N  \, \ln 2 \,,
\ee
that means that each of the $2^N$ configurations is accounted for in the same way, through $S = \ln$(number of microstates)$= \ln (2^N)$, see, e.g., \cite{kiefer}.

Notice that, within the model of \cite{aischol}, where the ultimate degrees of freedom, making both space and matter, and called $X$ons \cite{Xons}, the fermionic character gives precisely $\alpha = 4 \ln 2$, as shown in \cite{aismal}. According to different physical considerations, different choices of $\alpha$ are found in the literature \cite{Hod:1998vk,maggiore_2008}. Such quantization of the area is also an explicit result of Loop Quantum Gravity \cite{Ashtekar:1997yu,Rovelli:2004tv}.

This general picture was proposed by Mukhanov and Bekenstein long ago \cite{mukhanov,bekenstein2, bekenstein3}, and called ``the quantum BH''. That view made it till our days, although quantum corrections to the formula (\ref{ClassicalEntropy}), such as $S_{BH} \sim A/4 - \beta \ln A$, with $\beta > 0$, are often taken into account \cite{Kaul:2000kf, Gupta:2001bg, Ghosh:2004rq, Majhi:2013tw, Singleton:2013ama, Ong:2018xna}. Such corrections mean, of course, that not all configurations have the same weight. On this see, e.g., \cite{kiefer}. With this in mind, the study of the microscopic configurations of a BH amounts to the study of the configurations of a spin model, see, e.g. \cite{kiefer}.

The (quantum) information-theoretical approach to BH entropy/information evoked in the ``it from (qu)bit'' choice appears to clash with the latter, because if BH is the ultimate quantum computer, see, e.g., \cite{lloyd}, then one expects that: first, all configurations are treated equally, at the start of the computation; and second, the evolution of the quantum states through an Hamiltonian is not fixed by a given spin model, but rather governed by different Hamiltonians depending on the ``computation'' (for the factorization, see, e.g., \cite{Shor} and the semiclassical counterpart \cite{SemiShor}). Furthermore, certain arguments suggest that the fermionic degrees of freedom ($X$ons) are actually free in the BH phase, see, e.g., \cite{aischol} and \cite{aismal}, and this clashes with an interacting spin model that is necessary to explain why certain configurations dominate over others.

In \cite{Harms:1992nb,Harms:1992jt,Huang:2000kga} it was remarked that BHs, in the microcanonical description, fulfill a bootstrap condition \cite{hagedorn1,frautschi,nahm,hagedorn3} and should then exhibit self-similarity. That is, a BH could be viewed as formed by smaller BHs, formed in turn of even smaller BHs, and so on. Such picture becomes even more suggestive if we recall that the Hawking temperature \cite{haw1} of a Planck-sized BH ($T_H \approx l^{-1}_P$) could be viewed as the Hagedorn temperature in string theory \cite{veneziano1,susskind1,susskindBOOK}. At that temperature, BH evaporation stops and a phase transition is expected to occur, along the line of the transition between the hadronic phase and the quark-gluon-plasma phase \cite{Cabibbo:1975ig}. There, such behavior is described by a bootstrap equation \cite{hagedorn1,frautschi,nahm,hagedorn3}. Heavy hadronic resonances behave as if they were made of lighter resonances, and then the latter as if they were made of even lighter resonances, and so on, till the Hagedorn temperature.

To our knowledge, a precise mathematical description of that type was not proposed for BHs. Then, BH self-similarity is only a little-explored landscape. In this paper, mainly motivated by such considerations and not resorting to any specific spin-model, we propose a simple mathematical construction which explicitly describes BH self-similarity, within the quantum BH picture.

Let us recall that, the string Hagedorn temperature derives from the typical dual resonance model spectrum \cite{veneziano1}, $M^2 \simeq N$, where $N$ in the integer labeling the states. This can be seen by evaluating the number of \textit{unordered} partition of the integer (UPI) $N$, which grows as $\exp\{\gamma \sqrt{N}\}$, for some constant $\gamma$ \cite{barton}. An exponential mass spectrum can be obtained by a bootstrap equation, as for the hadronic resonances \cite{hagedorn1}, hence a similar equation should hold in the string description.

On the other hand, in the purely gravitational picture described earlier, the number of configurations one has to include into $S_{BH}$ grows as $2^N$. The relation of this number, $2^N$, with the number of \textit{ordered} partitions of the integer (OPI) $N$, that is $2^{N-1}$, is our main concern here. We would like to reconcile these two numbers, entirely within the quantum BH picture of \cite{mukhanov,bekenstein2, bekenstein3}, and then, knowing that self-similarity is at work in the OPI \cite{Blanchard:2004du}, move from there to extract from this connection some information on BH self-similarity, based entirely on the gravitational description.

The paper is organized as follows. In Section \ref{sectionSpinFlip} we construct a two-to-one map between the BH configurations and the OPI. In Section \ref{sectionNewOperation} we use that map to construct a composition law between the sub-parts making the whole configuration space that is in one-to-one correspondence with the OPI. We are then in the position to discuss how the self-similarity of OPI can find an incarnation in the self-similarity of the BH, as we do in Section \ref{sectionSelfSimilarity}. Finally, in Section \ref{sectionOnLogA} we compare the above to the well-known results on the subleading $\ln A$ corrections, and offer our conclusions in Section \ref{sectionConclusions}.


\section{The spin-flip map}\label{sectionSpinFlip}

What in the physics literature is often referred to as OPI of the (positive) integer $N$, in more precise mathematical terms, is usually called \textit{composition} of $N$ (where the UPIs are called \textit{decompositions}). Thus compositions are decompositions of $N$ as the sum of positive integers in which the order of the terms matters. The number of compositions of $N$, $O(N)$, is such that, if the first term of a composition of $N$ is $n$, the remaining terms are a composition of $N - n$, and there are $O(N-n)$ such compositions. This leads to
\be \label{selfsimOPI}
O(N) = 1 + O(N-1) + O(N-2) + \cdots + O(1) \,,
\ee
therefore $O(N+1) = O(N) + O(N-1) + \cdots + O(1) + 1 = O(N) + O(N) = 2 O(N)$, which is solved by $O(N) = 2^{N-1}$, consistently with $O(1)=1$. Notice that the recursive relation (\ref{selfsimOPI}) is an instance of the OPI self-similarity.

Thus, on the one hand, $O(N) = 2^{N - 1}$, whereas the number of configurations, $C(N)$, of $N$ copies of a two-level spin-$1/2$ system, for instance with one spin accommodated into each of the $N$ (Planck-sized) plaquettes making the area $A$, is given by $C(N) = 2^N$. Therefore, if we want to relate the two ways of counting configurations, we need to find a 2-to-1 map from the latter to the former.

Of course, there are many ways to do that, but the crucial point is to understand which spin configurations correspond to a specific OPI. The way we find here has many advantages, especially the fact that the self-similarity will then be obtained in a simple and consistent way.

As will be clarified later, by mapping the OPI of N to the corresponding spin configurations, we need to distinguish between configurations differing not only  by how many spin are up and how many are down, but also by the \textit{position} of the spin. E.g., the configuration $(\uparrow, \uparrow, \downarrow)$ needs to be distinct from the configuration $(\uparrow, \downarrow, \uparrow)$ and from the configuration $(\downarrow, \uparrow, \uparrow)$. Other countings do not distinguish between those configurations, hence carry a degeneracy that is only a function of how many arrows are down (one here) and how many arrows are up (two here), see, e.g., \cite{kiefer}.

When we distinguish, we indeed have $2^N$ configurations, that is precisely what we need to explain the result in equation (\ref{ClassicalEntropy}), i.e., $S = \ln$(number of microstates)$= \ln (2^N)$. Now we want to find how this number is related to the number of ways one can make an OPI of $N$, that is known to be $2^{N-1}$. Thus, we need to halve the number of BH configurations, and we need to do that in a consistent way, in order to associate spin states, on the one hand, with partitions of $N$, on the other hand.

One way to do that is to identify the BH configurations that are related by a full ``spin-flip'', that is: to any \textbf{one} given OPI of $N$ we associate \textbf{two} BH states that can be obtained one from the other when \textit{all} the spins that identify the given configuration change their orientation, $\uparrow \leftrightarrow \downarrow$. For instance, a particular partition of $N=3$ (1+1+1, in this case, see later) we make it correspond to $(\uparrow, \downarrow, \uparrow)$ \textit{and} to $(\downarrow, \uparrow, \downarrow)$.

This identification is motivated, \textit{a posteriori}, by the fact that later we shall be able to propose a composition law among sub-systems of the configuration space, consistently related to the integers composing $N$, that gives back precisely the configuration space we started with and this will enable us to discuss self-similarity.

The other ingredient of the map we are constructing is a rule that relates a given pair of BH (spin) configurations to a given OPI. We found that the following rule does the job.

\textit{When a spin is next to an opposite spin, i.e., when $\uparrow$ is next to $\downarrow$ or when $\downarrow$ is next to $\uparrow$, in the OPI this corresponds to 1 + 1. E.g., $(\uparrow, \downarrow, \uparrow, ...)$ and the spin-flipped $(\downarrow, \uparrow, \downarrow, ...)$ both correspond in the OPI to the partition $1+1+ ...$. When the spin are likewise they contribute with an integer that is the sum of how many times the spin does not flip. E.g., $(\uparrow, \uparrow, \downarrow, ...)$ and $(\downarrow, \downarrow, \uparrow, ...)$ correspond in the OPI to the partition $2 + ...$.}

Together with the spin-flip, with this rule one takes into account all possibilities. These make the wanted two-to-one map from the BH configurations to the OPI. We call it the ``spin-flip map''.

As an example, let us write the spin configurations corresponding, through the spin-flip map, to a specific partition of $N=7$:
\be
1+3+2+1 \to
  \begin{array}{ccccccc}
    (\uparrow,   & \downarrow, & \downarrow, & \downarrow, & \uparrow,   & \uparrow,   & \downarrow) \\
    (\downarrow, & \uparrow,   & \uparrow,   & \uparrow,   & \downarrow, & \downarrow, & \uparrow) \\
  \end{array} \,. \nonumber
\ee

Let us also show how the spin-flip map works for all the OPI of $N=4$
\bea
1+1+1+1 &\to& (\uparrow, \downarrow, \uparrow, \downarrow)   \nonumber \\
2+1+1   &\to& (\uparrow, \uparrow, \downarrow, \uparrow)     \nonumber \\
1+2+1   &\to& (\uparrow, \downarrow, \downarrow, \uparrow)   \nonumber \\
1+1+2   &\to& (\uparrow, \downarrow, \uparrow, \uparrow)     \nonumber \\
2+2     &\to& (\uparrow, \uparrow, \downarrow, \downarrow)   \nonumber \\
3+1     &\to& (\uparrow, \uparrow, \uparrow, \downarrow)     \nonumber \\
1+3     &\to& (\uparrow, \downarrow, \downarrow, \downarrow) \nonumber \\
4       &\to& (\uparrow, \uparrow, \uparrow, \uparrow)  \label{N=4}    \,.
\eea
The number of OPI $O(N)$ we obtain is precisely $8 = 2^3 = O(4)$, while, since the spin-flip is understood everywhere, on the right side we count $16 = 2^4 = C(4)$ configurations.

Notice that in \cite{mukhanov}, the multiplicity $2^N$ has been interpreted as the number of ways in which a BH, in the $N$th area level, can be made by first making a BH in the ground state, and then going up in the ladder of all possible levels. It has been also discussed in \cite{danielsson} that such multiplicity may rather represent the number of ways in which the BH  can ``decay'' down the staircase of levels to the ground state. For both
interpretations see also \cite{bekenstein2} and \cite{bekenstein3}.

However, in those papers, no specific correlation has been discussed with the spin configurations. Here, instead, we do so, even though we do not commit ourselves to any specific physical model that would give meaning to the spin-flip map. Actually the spin-flip symmetry (degeneracy) may even be solely instrumental to the construction of the map with the OPI and it should be as well lifted-up, when it comes to compute the actual information content of a given configuration. If we do so, then a proper information-theory approach becomes possible, giving meaning to the ``it for qubit'' choice discussed earlier. Indeed, the possibility to distinguish among configurations that have the same number of $\downarrow$s and of $\uparrow$s, but arranged in different positions in the string, e.g. $(\downarrow, \downarrow, \uparrow) \neq (\uparrow, \downarrow, \downarrow)$, allows such information-theoretical approach, where the position in the string of values is crucial. E.g., for the example above, we are saying that 1 is different from 8, in binary notation, $001 \neq 100$ (we simply use $\uparrow \equiv 1$ and $\downarrow \equiv 0$).

\section{The composition law}\label{sectionNewOperation}

Having established the spin-flip map, between BH configurations and OPIs, we want now to see how the self-similarity patterns of the OPI can be imported/translated into the self-similarity of BHs.

This is less easy than it might seem, because we need to find a way to construct the given configuration space, of dimension $2^N$ starting from \textit{any} combination of the smaller configurations spaces, of dimensions $2^{N_i}$s, with $\sum_i N_i = N$, but, at the same time, we must keep the correspondence with the partition of integers.

To start, let us notice that, in the example (\ref{N=4}) we can read the partition of the smaller number encoded into the lager, here the partition of $N=3$ into that of $N=4$
\bea
1+1+1+1 &\to& (\underbrace{\uparrow, \downarrow, \uparrow}, \overbrace{\downarrow})    \to (1+1+1)+1   \nonumber \\
2+1+1   &\to& (\underbrace{\uparrow, \uparrow, \downarrow}, \overbrace{\uparrow})     \to  (2+1)+1    \nonumber \\
1+2+1   &\to& (\underbrace{\uparrow, \downarrow, \downarrow}, \overbrace{\uparrow})   \to  (1+2)+1    \nonumber \\
1+1+2   &\to& (\overbrace{\uparrow}, \underbrace{\downarrow, \uparrow, \uparrow})     \to  1+(1+2)    \nonumber \\
2+2     &\to& (\underbrace{\uparrow, \uparrow}, \underbrace{\downarrow, \downarrow})   \to  (2)+(2)    \nonumber \\
3+1     &\to& (\underbrace{\uparrow, \uparrow, \uparrow}, \overbrace{\downarrow})     \to  (3)+1      \nonumber \\
1+3     &\to& (\overbrace{\uparrow}, \underbrace{\downarrow, \downarrow, \downarrow}) \to  1+(3)      \nonumber \\
4       &\to& (\uparrow, \uparrow, \uparrow, \uparrow)  \quad {\rm NEW}                  \label{N=4 and N=3}\,.
\eea
On the right side we read in parentheses the partitions of $N=3$, obtained on the left side by grouping spins with the underbrace. That is indeed a way to extract a smaller partition from a bigger partition, but it is easy to see that there is some unwanted degeneracy on the right side. This is evident in the fifth line, where we have twice the partition of 2, but it is also true that we could have grouped the spins of the first line to obtain $1 + (1+1+1)$, or the spins of the third line to obtain $1+(2+1)$, etc. The degeneracies, clearly, grow when we include the partitions of the next smaller number, 2 here, and, in general, they grow with $N$. So, we can indeed trace the partitions of the smaller numbers within the partitions of the given number $N$, but we have a problem with a heavy over-counting.

To avoid that, we shall work with the spin configurations, and define an operation, $\hat{+}$, that allows to obtain the configuration space of the given BH made of $N$ plaquettes, of dimension $C(N) = 2^N$, \textit{only once} for any given partition. In other words, if we indicate with $\mathbf{N}$ such configuration space, and $N_1 + N_2 + \cdots = N$ is a given OPI of N, we want that $\mathbf{N_1} \hat{+} \mathbf{N_2} \hat{+} \cdots = \mathbf{N}$. Doing so we would establish a one-to-one correspondence between the OPI of $N$, and the way to combine the subspaces of $\mathbf{N}$, corresponding to the OPI. The operation is constructed as follows:


\textit{We first list all the OPIs of $N$. Then take each partition in the list, say $N_1 + N_2 = N$, and write the spin configurations for $\mathbf{N_1}$, the configuration space associated to the first number of the sum. (To simplify the operation we can consider even just one representative per each spin-flipped pair, even though, to obtain all the configurations, at the end of the procedure the spin-flip needs be applied). Then, we take the tensor product of each of those representatives with \textbf{all} the spin configurations of $\mathbf{N_2}$ \textbf{explicitly including} the spin-flipped configurations. The result of such operation, $\mathbf{N_1} \hat{+} \mathbf{N_2}$, are all the spin configurations of $\mathbf{N}$, with no redundant nor missed configuration. (Clearly, if the spin-flip was understood at the start, then so will be in the final result). The operation will give the same result for each OPI of $N$, including those with more than two terms. For the latter, one must start from the first term on the left, act with the second as just described, and the result of this needs be acted upon with the next term, and so on, till the end.}

 The trivial example is $\mathbf{N}=\mathbf{N}$, where no composition is performed. The first non-trivial operation is $\mathbf{1} \hat{+} \mathbf{1}$, that originates from the partition 1+1=2, so it must give $\mathbf{2}$:
\be
\mathbf{1} \hat{+} \mathbf{1} = \uparrow \otimes \begin{array}{c}
                 \rc{\uparrow} \\
                 \rc{\downarrow}
               \end{array} =
               \begin{array}{c}
               \uparrow \rc{\uparrow} \\
               \uparrow  \rc{\downarrow}
               \end{array}
               = \mathbf{2} \,.
\ee
Here, in the second-last term, the first line is one spin representative of 2, $(\uparrow, \uparrow)$, while the second line is one spin representative of 1+1, $(\uparrow, \downarrow)$. As said, the four-dimensional ($2^N=2^2$), full configuration space, $\mathbf{2}$, is obtained when we spin-flip each final configuration: $(\uparrow, \uparrow)$,$(\downarrow, \downarrow)$ and $(\uparrow, \downarrow)$,$(\downarrow, \uparrow)$.

Let us now show how to act with three addends, e.g., $\mathbf{1} \hat{+} \mathbf{1} \hat{+} \mathbf{1}$ that is one of the four ways ($2^{N-1} = 2^2$) to obtain $\mathbf{3}$
\be
(\mathbf{1} \hat{+} \mathbf{1}) \hat{+} \mathbf{1} = \left( \uparrow \otimes \begin{array}{c}
                 \rc{\uparrow} \\
                 \rc{\downarrow}
               \end{array} \right) \hat{+} \mathbf{1}
               =
               \left( \begin{array}{c}
               \uparrow \uparrow \\
               \uparrow \downarrow
               \end{array} \right) \otimes  \begin{array}{c}
                 \rc{\uparrow} \\
                 \rc{\downarrow}
               \end{array} = \begin{array}{c}
               \uparrow \uparrow \rc{\uparrow}\\
               \uparrow \uparrow \rc{\downarrow}\\
               \uparrow \downarrow \rc{\uparrow}\\
               \uparrow \downarrow \rc{\downarrow}
               \end{array}
               = \mathbf{3} \,.
\ee
Here, in the second-last term, the first line is the spin representative of 3, $(\uparrow, \uparrow, \uparrow)$, the second line is the spin representative of 2+1, $(\uparrow, \uparrow, \downarrow)$, the third line is the spin representative of 1+1+1, $(\uparrow, \downarrow, \uparrow)$, and the third line is the spin representative of 1+2, $(\uparrow, \downarrow, \downarrow)$. Again, the full configuration space $\mathbf{3}$, of dimension eight, is obtained when we spin-flip each of the four final configuration.

It is crucial to notice again that this operation gives $\mathbf{N}$, the full configuration space, of dimension $2^N$, in a non degenerate way, each time we use any of the $2^{N-1}$ OPIs of $N$. Therefore, the 8-dimensional $\mathbf{3}$ ($8 = 2^3$) is obtained 4 times ($4=2^{3-1}$) as $\mathbf{1} \hat{+} \mathbf{1} \hat{+} \mathbf{1}$, $\mathbf{2} \hat{+} \mathbf{1}$, $\mathbf{1} \hat{+} \mathbf{2}$ and $\mathbf{3}$.

To show how $\hat{+}$ acts when none of the addends is $\mathbf{1}$ let us show the case $\mathbf{2} \hat{+} \mathbf{2}$ that should give $\mathbf{4}$
\be
\mathbf{2} \hat{+} \mathbf{2} =
               \begin{array}{c}
               \uparrow \downarrow \\
               \uparrow \uparrow
               \end{array}
               \otimes
               \begin{array}{c}
               \rc{\uparrow \downarrow} \\
               \rc{\downarrow \uparrow} \\
               \rc{\uparrow \uparrow}   \\
               \rc{\downarrow \downarrow}
               \end{array}
               =
               \begin{array}{c}
               \uparrow \downarrow  \rc{\uparrow \downarrow} \\
               \uparrow \downarrow  \rc{\downarrow \uparrow} \\
               \uparrow \downarrow  \rc{\uparrow \uparrow}   \\
               \uparrow \downarrow  \rc{\downarrow \downarrow} \\
               \uparrow \uparrow \rc{\uparrow \downarrow} \\
               \uparrow \uparrow \rc{\downarrow \uparrow} \\
               \uparrow \uparrow \rc{\uparrow \uparrow}   \\
               \uparrow \uparrow \rc{\downarrow \downarrow}
               \end{array} =
               \mathbf{4} \,.
\ee
Here, in the second-last term, scrolling from the first to the last line,  we recognise one spin representative of: 1+1+1+1, 1+2+1, 1+1+2, 1+3, 3+1, 2+1+1, 4 and 2+2, i.e, of all the 8 OPIs of 4. As said, the full 16-dimensional $\mathbf{4}$ is obtained via spin-flip.

Once we have exploited the first simple cases, the general proof follows by induction. In fact, suppose the above composition rule holds for $\mathbf{N}$. Then
\be
  \mathbf{1} \, \hat{+} \,  \mathbf{N} \ = \ \mathbf{1}\,\hat{+} \, \mathbf{N_1} \, \hat{+} \, \mathbf{N_2} \, \hat{+} \,  \cdots \, \hat{+} \mathbf{N_M}    \, ,
\ee
where $N_1 + N_2 + \cdots N_M=N$, with $M \leq N$. Because $N_1 \leq N$, we can apply the composition rule to get
\be
\mathbf{1} \, \hat{+} \,  \mathbf{N}  \ = \ \left(\mathbf{N_1+1}\right) \, \hat{+} \, \mathbf{N_2} \, \hat{+} \,  \cdots \, \hat{+} \mathbf{N_M} \ = \ \mathbf{N+1} \, ,
\ee
which concludes our proof.
\section{The self-similarity}\label{sectionSelfSimilarity}

If the configuration space $\mathbf{N}$ is that of a quantum BH \textit{a la} Bekenstein \cite{bekenstein3}, then we have found a clean and consistent way to see BH self-similarity:

\textit{Keeping the BH fixed, its configuration space $\mathbf{N}$ is made of the configuration spaces of smaller BHs, that are made of configuration spaces of even smaller BHs, and again and again, until we reach $N$ copies of $\mathbf{1}$, the configuration space of the tiniest (elementary) BH.}

That is, since to any of the $2^{N-1}$ OPIs of $N$ we can associate one of the $2^{N-1}$ OPIs of $\mathbf{N}$
\be
\sum_{i} N_i \ =  \ N \to \hat{\sum}_{i} \mathbf{N_i} = \mathbf{N}  \,, \,\, \sum_{j} M_j = N \to \hat{\sum}_{j} \mathbf{M_j} = \mathbf{N} \,, ... \,,
\ee
where $\hat{\sum}_{i} \mathbf{N_i} = \mathbf{N_1} \hat{+} \mathbf{N_2} \hat{+} \cdots $, whatever pattern we found in the OPI of $N$, it is found in the configuration space $\mathbf{N}$ of the BH, and then repeated for the smaller numbers, until we reach the ``quantum'' of the BH space, $\mathbf{1}$.

For instance, a suggestive pattern is given by
\be
\mathbf{N} \ = \ \mathbf{1} \hat{+} (\mathbf{N-1}) \ = \ \mathbf{1} \hat{+} (\mathbf{1} \hat{+}  (\mathbf{N-2})) = \mathbf{1} \hat{+} (\mathbf{1} \hat{+} (\mathbf{1} \hat{+}  (\mathbf{N-3}))) = \cdots
= \hat{\sum}^{N}_{i=1} \mathbf{1} \,.
\ee
Here one can say that when the configuration space of the tiniest BH, $\mathbf{1}$, is isolated from the rest, than this can be repeated again and again till the complete splitting.

One crucial comment is that, as wanted, in this picture the self-similarity does not require any change of description of the degrees of freedom. We keep the BH fixed and we are just finding patterns within the configuration space of that given BH. In other words, we do not deal here neither with BH \textit{evaporation} \cite{haw1} nor we deal with BHs \textit{merging} \cite{abbotts}.

In the first case, we have a configuration space that ``decays'' till complete evaporation \cite{aismal}: $\mathbf{N} \to \mathbf{N-1} \to \cdots \to \mathbf{0}$, and the self-similarity is something that should emerge from the Hagedorn spectrum, as discussed in \cite{susskind1,veneziano1}, but that was never explicitly found, e.g., by solving the related bootstrap equation. An approximate self-similarity in BH evaporation in the semiclassical limit was also discussed in Ref. \cite{Dvali:2012gb}.

In the second case, when we merge two BHs, one of area $A_1$ and mass $M_1$, with $A_1 = \alpha N_1 \sim M_1^2$, one of area $A_2$ and mass $M_2$, with $A_2 = \alpha N_2 \sim M_2^2$, and with $\alpha = 4 l_P^2  \ln 2 $, the area of the resulting BH of mass $M = M_1 + M_2$ is
$A \sim (M_1 + M_2)^2$, i.e.,
\be
A = \alpha N = \alpha (N_1 + N_2 + 2 \sqrt{N_1 N_2}) \,,
\ee
which points to
\be
S \ = \ S_1 \ +\  S_2 \ + \  2 \sqrt{S_1 S_2} \,.
\ee
In other words, the configuration spaces $\mathbf{N_1}$ and $\mathbf{N_2}$, of dimensions $2^{N_1}$ and $2^{N_2}$, respectively, when combined through the operation that we have constructed, $\hat{+}$, give a configuration space of dimension $2^{N_1 + N_2}$, which is much smaller than the configuration space of the BH obtained by merging the two BHs, that is
\be
2^N \ = \  2^{N_1 + N_2} \times \left( 2^{\sqrt{N_1 N_2}} \right)^2 \,.
\ee

\section{Configurations Analysis}\label{sectionOnLogA}


In the previous Sections, consistently with a quantum information-theoretical approach and with the picture of a BH as a free (non-interacting) gas of fermions ($X$ons), we have treated all configurations in $\mathbf{N}$ equally. Let us compare here our analysis to the literature \cite{kiefer}, where interesting questions are posed on weighting differently the different configurations, and how this can be related to the subleading corrections to the Bekenstein formula, $S_ {BH} \sim A/4 - \beta \ln A$. Such corrections are usually referred to as ``quantum corrections'', as indeed they are obtained in perturbative quantum theories \cite{Kaul:2000kf, Gupta:2001bg, Ghosh:2004rq, Majhi:2013tw, Singleton:2013ama, Ong:2018xna}. Here we do not call them so, because we are within a quantum BH model \cite{mukhanov,bekenstein2,bekenstein3}, in the first place.

As discussed, in our case the actual position of the spin within the given configuration matters, so that, e.g., $8 \, (= 100) \neq 1 \, (= 001)$, where in parentheses we indicate the binary number. This is not the point of view of interacting spin systems, taken elsewhere \cite{kiefer}. There, in general, the various configurations, characterized by how many spins are up and how many are down, are labeled by the energy.

However, to reproduce the BH entropy in that approach, one cannot select a unique configuration. A large entropy requires that the lowest energy state corresponds to a large subset of the possible $2^N$ configurations. Therefore, in that picture, the spin-spin coupling between spins in positions $i$  and $j$, say it $J_{i,j}$, should actually be insensitive to such positions, i.e., $J_{i,j} \equiv J$.  For example, one can consider the Ising model in an antiferromagnetic state, with constant spin-spin coupling, $J<0$, and energy $E$.

The energy levels of such system do not depend on the position of the spin. If we have $N$ total spins, and $n$ of those are flipped with respect to the remaining $N-n$, the energy levels are given by  \cite{Czachor1}
\be \label{eq17}
E_n\ = \ \frac{J}{4}[n-\frac{1}{2}(N-n/2)(N-n/2-1)] \,,
\ee
where, for simplicity we take $N$ even and where $n=0,1,2,..N$. The total number of energy levels is $N+1$.

The degeneracy of level $n$ is
\be
d_n=B(N,n) \ = \  \frac{N!}{n!(N-n)!} \,.
\ee

The lowest energy level is for $n=N/2$, with degeneracy
\be
d_{N/2} \ = \ \frac{N!}{[(N/2)!]^2}\simeq \frac{2^N}{\sqrt{N}} \sqrt{2/\pi} \,,
\ee
where the Stirling approximation has been used. Therefore, the resulting entropy of this particular subset of the whole set of configurations is
\be
S(N/2) \ = \ N \ln 2 - \frac{1}{2} \ln N + constant \,,
\ee
which gives a logarithmic subleading correction, as calculated with various approaches with different coefficients of the $\ln N$ term \cite{Kaul:2000kf, Gupta:2001bg, Ghosh:2004rq, Majhi:2013tw, Singleton:2013ama, Ong:2018xna}. This may be considered a point in favour of this approach, but it also has evident drawbacks.

First, this state has negative energy $E = J N/8 <0$ proportional to the total number of spins, i.e. to the number of Planck plaquettes, $N$, whereas the BH mass is proportional to $\sqrt N$.

Second, by an \textit{ad hoc} choice of the coupling, $J \simeq 1/\sqrt{N}$, a \textit{fine-tuning}  with the large amount of positive energy is required to get the BH mass. This  problem can be overcome only by assuming the specific coupling $J=K/N$, i.e the Kac model \cite{Kac}, since in this case the BH mass should be \cite{Czachor1}
\be \label{eq21}
M \ = \  M_{P} \sqrt N \, + \, E_{N/2} \ = \  M_{P} \sqrt N \, + \, K/8 \,.
\ee

Finally, the coefficient of the $\ln N$ correction crucially depends on the specific configuration, henceforth, if some other configuration contributes too, the coefficient changes accordingly. Indeed the total degeneracy in the general case is given by
\begin{equation}
D(E) \ = \ \Sigma_{n \ne N/2} \frac{B(N,n)}{|n-N/2|} [\delta(n-n_{+}(E))+\delta(n-n_{-}(E))] \, + \, d_{N/2} \delta(E-E_{N/2}) \,,
\end{equation}
where
\be
n_\pm \ = \ \frac{1}{2} [N \pm \sqrt{(N-8E/J)}]
\ee
Let us consider a configuration with $n=N/2 + m$ with $m<<N/2$. From the previous equation one obtains
\be
m \ = \ \frac{1}{2} [\sqrt{(N-8E/J)}]
\ee
The corresponding term in the degeneracy is
\be
d_m  \ = \ \frac{1}{m} B(N,N/2+m)
\ee
which, after some algebra by using  Stirling formula, turns out to be
\be
d_m \ \simeq \  2^N \frac{1}{m} \frac{1}{\sqrt{2\pi N}} \frac{1}{(1-4m^2/N^2)^{N/2+1}} \frac{1}{1-m^2/2N^2}
\ee
For large $N$, the most relevant correction to the entropy associated to $d_{N/2}$ comes from the $1/m$ term ($c$ and $c'$ being numerical constants or $O(1/N)$ terms) by Eqs.(\ref{eq17})-(\ref{eq21}), i.e.
\be
S(m) \ \simeq \ N \ln 2 - \frac{1}{2} \ln N - \ln m + c =  N \ln 2 - \ln N + c'
\ee
with a change in the coefficient of the $\ln N$ correction, that, for large $N$ does not depend on the different choices of the coupling $J$ ( $J \sim$constant, $J \sim O(1/\sqrt{N})$ or $J \sim O(1/N)$). Therefore, the coefficient of the subleading contributions originates from a specific, although large, subset of all the possible configurations.

\section{Conclusions}\label{sectionConclusions}

Within a quantum model of BHs, whose associated degrees of freedom are finite in number and fermionic, we have studied general aspects of its entropy. In particular, we have established a precise correspondence between the OPI of a natural number $N$ and the configuration (Hilbert) space of the BH, of dimension $2^N$. This is done by constructing a two-to-one map from the former to the latter, and then a composition law, $\hat +$, based on the tensor product, between sub-parts of the configuration space.

Through this composition law, we have been able to account for BH self-similarity through the well-known self-similarity of OPI. In our approach, such self-similar structure of the BH's configuration space emerges in a transparent way, and the operation $\hat +$ allows to always obtain all possible $2^N$ configurations.

The point of view taken here considers non-interacting (free) fermionic degrees of freedom (spins), along the lines of \cite{aischol,Xons,aismal}, with equally probable configurations, identified by number and position of the spin in the (information theoretical) \textit{string}.

If we want to use an information-theoretical approach, the selection of a specific configuration over the others is not appropriate. To use the BH as the ultimate quantum computer \cite{lloyd}, then one expects that all configurations are treated equally, and the evolution of the quantum states should not be fixed by a given spin model, but should rather be governed by a specific Hamiltonian that ``implements'' the given ``computation''.

\section*{Acknowledgements}
P.C. and A.I. gladly acknowledge support from Charles University Research Center (UNCE/SCI/013). L.S. was supported by the Polish National Science Center grant 2018/31/D/ST2/02048.

%





%
%
%


\end{document}